# Edge AI Deployment Beyond Models: A BSP-Aware Systems Framework for Industrial Embedded Platforms

**Pitchai Muthu M**

*Embedded & IoT Software Team Lead*
Advantech Industrial Computing India Pvt. Ltd., India
Pitchai.muthu@advantech.com

## ABSTRACT

*Industrial Edge AI programs often begin with the model and only later confront the platform. That sequencing is attractive because it allows early demonstrations, but it breaks down when the deployment target is an embedded system with long product lifecycles, vendor-specific kernels, heterogeneous accelerators, safety constraints, and nontrivial I/O paths. In that environment, a model is only one component of a larger execution chain that begins at the sensor, traverses the board support package (BSP), and ends in a production service loop. This paper argues that robust Edge AI deployment must be treated as a systems problem rather than a late-stage application packaging exercise.*

*The paper presents a BSP-aware framework for industrial embedded platforms organized around five layers: hardware, BSP/operating-system adaptation, runtime and acceleration, application/inference, and operations/validation. The discussion is grounded in vendor architecture documentation for Android, NXP i.MX, NVIDIA Jetson, ONNX Runtime, and TensorRT, and in systems literature on embedded AI benchmarking, device instability, and heterogeneous edge fleets. The result is a practical framework that connects low-level platform work to measurable deployment outcomes such as reproducibility, diagnosability, sustained throughput, and field reliability.*

## EXECUTIVE SUMMARY

Industrial Edge AI deployments are increasingly constrained not by model performance but by system-level integration challenges arising from Board Support Package (BSP), driver, and hardware dependencies. Conventional model-first deployment approaches assume abstraction from underlying system layers; however, in industrial embedded environments, such assumptions are invalid. Failures frequently originate from kernel subsystems, peripheral drivers, and hardware interfaces, leading to unstable inference behavior, delayed deployments, and increased debugging overhead.

To address these limitations, this work proposes a BSP-aware, systems-first deployment framework structured around a five-layer architecture comprising hardware, BSP, operating system, middleware, and application layers. The framework introduces a gated lifecycle methodology (Phases 1–5 with validation Gates A–D), enforcing progressive verification of system readiness prior to AI integration. This approach ensures that each layer is functionally validated and performance-aligned before advancing to subsequent stages, thereby minimizing cascading failures.

Key findings indicate that BSP and lower system layers form an integral part of the effective inference path, directly influencing data acquisition, processing latency, and output reliability. The adoption of a gated, systems-first workflow reduces late-stage integration risks, improves debugging efficiency, and enhances deployment predictability across heterogeneous hardware platforms. While this methodology may introduce additional upfront validation, effort compared to rapid model-first prototyping, it delivers significant gains in system robustness, scalability, and long-term maintainability, particularly in mixed-fleet industrial deployments.

# I. INTRODUCTION

## A. Why This Topic Matters Now

Edge AI has become a default architectural choice for many industrial systems because inference at the device reduces round-trip latency, lowers backhaul dependence, and keeps sensitive operational data closer to the machine that generated it. The technical conversation, however, is still disproportionately focused on model conversion, quantization, and framework choice. Those are important topics, but they do not explain why a model that behaves correctly on a workstation or on a development kit can become unstable, slow, or operationally expensive once it is placed inside a real industrial product.

Industrial embedded platforms differ from generic edge nodes in several ways. They are usually expected to remain in service for years, to tolerate thermal and power variation, to work with mixed I/O topologies, and to support service procedures such as rollback, recovery, and reproducible field diagnostics. In such systems, low-level integration is not a background detail. The platform stack itself determines whether the model receives correctly timed inputs, whether the accelerator is actually exercised, whether memory pressure stays within bounds, and whether the device can be updated without bricking the fleet. The BSP is therefore not an implementation footnote; it is part of the deployment boundary.

## B. The Central Problem

A large fraction of Edge AI deployment failures are not caused by an incorrect model architecture. They are caused by system behavior that changes the conditions under which the model runs: camera pipelines that alter image characteristics, audio paths that drift from the expected sample format, thermal policies that reduce sustained throughput, or vendor-driver mismatches that silently disable the intended acceleration path. Android's architecture documentation explicitly places HALs and the kernel between higher-level system components and the hardware, emphasizing that hardware access is mediated through well-defined lower layers rather than accessed directly by the application [1]. NXP's i.MX Linux documentation similarly describes the BSP as the software base used to create the bootloader, kernel image, and root filesystem for the target platform [3]. NVIDIA's Jetson Linux documentation states that the BSP sits at the core of the stack and includes the kernel, bootloader, sample root filesystem, toolchain, and sources [4]. These are not optional details; they define the executable system.

## C. Thesis and Contributions

The thesis of this paper is straightforward: industrial Edge AI should be deployed with a systems mindset, and that mindset should be organized around BSP awareness from the beginning of the project. The contribution of the paper is not a new benchmark or a proprietary vendor claim. Instead, it provides a structured framework that translates the evidence base from vendor architecture documents and systems literature into an operational method for real deployments. Specifically, the paper offers four contributions:

1. It clarifies what "BSP-aware" means in industrial Edge AI practice and why that perspective is materially different from model-first deployment. It introduces a layered framework that ties hardware, BSP, runtime, application logic, and operational validation into one deployment model.
2. It identifies recurrent cross-layer failure modes and maps them to concrete diagnostic artifacts and mitigation actions.
3. It provides an evidence-based comparison between an application-first workflow and a systems-first workflow, without inventing unsupported performance numbers.

4. It maps the framework to representative industrial use cases and provides a production sign-off checklist for engineering teams.

### D. Scope and Method

This paper is written in the style of an industrial research white paper using an IEEE-like format. The analysis synthesizes public technical documentation and peer-reviewed systems research. It does not claim a new experimental benchmark dataset. When the paper uses the phrase "observed result," it refers to observations reported in the cited literature or in vendor documentation and then interpreted in the context of deployment engineering. That distinction matters because industrial white papers lose credibility when they overstate certainty or imply measurements that were never actually performed.

## II. WHY BSP AWARENESS MATTERS IN INDUSTRIAL EDGE AI

### A. What a BSP Is in Deployment Terms

In embedded engineering, the BSP is the platform-specific software layer that brings up the board and exposes hardware to the operating system and upper software stack. Depending on the platform, it includes the boot chain, kernel, device tree or equivalent hardware description, peripheral drivers, firmware, vendor libraries, configuration files, and often the reference root filesystem used for system construction. NXP's i.MX Linux User's Guide describes the BSP as the collection of binary files, source code, and support files used to create the U-Boot bootloader, Linux kernel image, and root filesystem [3]. NVIDIA's Jetson documentation makes the same point with slightly different packaging, locating the BSP at the center of the Jetson Linux stack [4]. Android's device architecture makes the platform boundary explicit through HALs and kernel integration points [1], [2]. Across vendors, the vocabulary differs, but the structural role is the same.

For Edge AI, that means the BSP determines whether the hardware interface seen by the model is stable, reproducible, and production-ready. The relevant questions are not limited to "does the board boot?" They include: are timestamps consistent, are DMA paths configured correctly, are sensor clocks stable, is the intended accelerator accessible through the chosen runtime, are memory carve-outs appropriate, and can the platform recover cleanly after update or power interruption? All of these sit below the application layer, but all of them influence AI behavior in production.

### B. Why a Model Can Fail Even When the Model Is Correct

A common misconception in deployment planning is that once the model passes offline validation, the remaining work is "integration." In industrial systems, that separation is artificial. The input that the model consumes is not abstract data; it is the output of a sensor path and a software path. Compression, ISP tuning, sample-rate conversion, buffering strategy, and synchronization policy all shape the final tensor. Cidon et al. found that model prediction divergence across mobile edge devices is materially affected by compression formats and image signal processing, and that 14–17% of images in their experiments produced divergent classifications across one or more device models [8]. Their work focused on mobile devices rather than industrial cameras, but the systems lesson carries directly into machine vision deployments: if the input pipeline is not treated as part of the model execution environment, "accuracy" measured elsewhere is not enough.

The same reasoning applies to audio and time-series systems. If the codec path clips peaks, the DMA cadence introduces jitter, or the runtime converts sample formats implicitly, the deployed signal distribution no longer matches what was validated in development. This is not an argument against model optimization; it is an argument that optimization must be constrained by the realities of the platform stack.

### C. Why Industrial Environments Amplify the BSP Problem

Industrial platforms are frequently deployed in mixed fleets and retained in the field for long service windows. The implication is that hardware heterogeneity is not a temporary inconvenience; it is part of the operating environment. Woisetschläger et al. note that edge systems are characterized by diverse hardware, energy constraints, and unreliable clients, and observe that embedded devices in current systems can be more than five years old when high reliability requirements are involved [10]. That point matters for Edge AI because deployment logic that assumes a single homogeneous target quickly becomes expensive once the installed base spans multiple board revisions, accelerator capabilities, or BSP release trains.

Industrial conditions also add requirements that many demos ignore: long-duration thermal behavior, restart determinism, service-port access, field update safety, boot-time guarantees, and traceability of software provenance. In other words, industrial Edge AI is not simply "AI on a small computer." It is AI embedded inside a product system with lifecycle responsibilities. A BSP-aware mindset acknowledges that reality early enough for architecture choices to remain affordable.

## III. A BSP-AWARE FRAMEWORK FOR INDUSTRIAL EMBEDDED PLATFORMS

### A. Framework Overview

The proposed framework is organized as five layers, shown in Figure 1 below. The value of this representation is that it turns a vague recommendation — "consider the system" — into explicit engineering boundaries. Each layer has distinct responsibilities, deliverables, and failure modes, but none of the layers can be validated in isolation for long. The system only becomes deployable when the interfaces between the layers are stable.

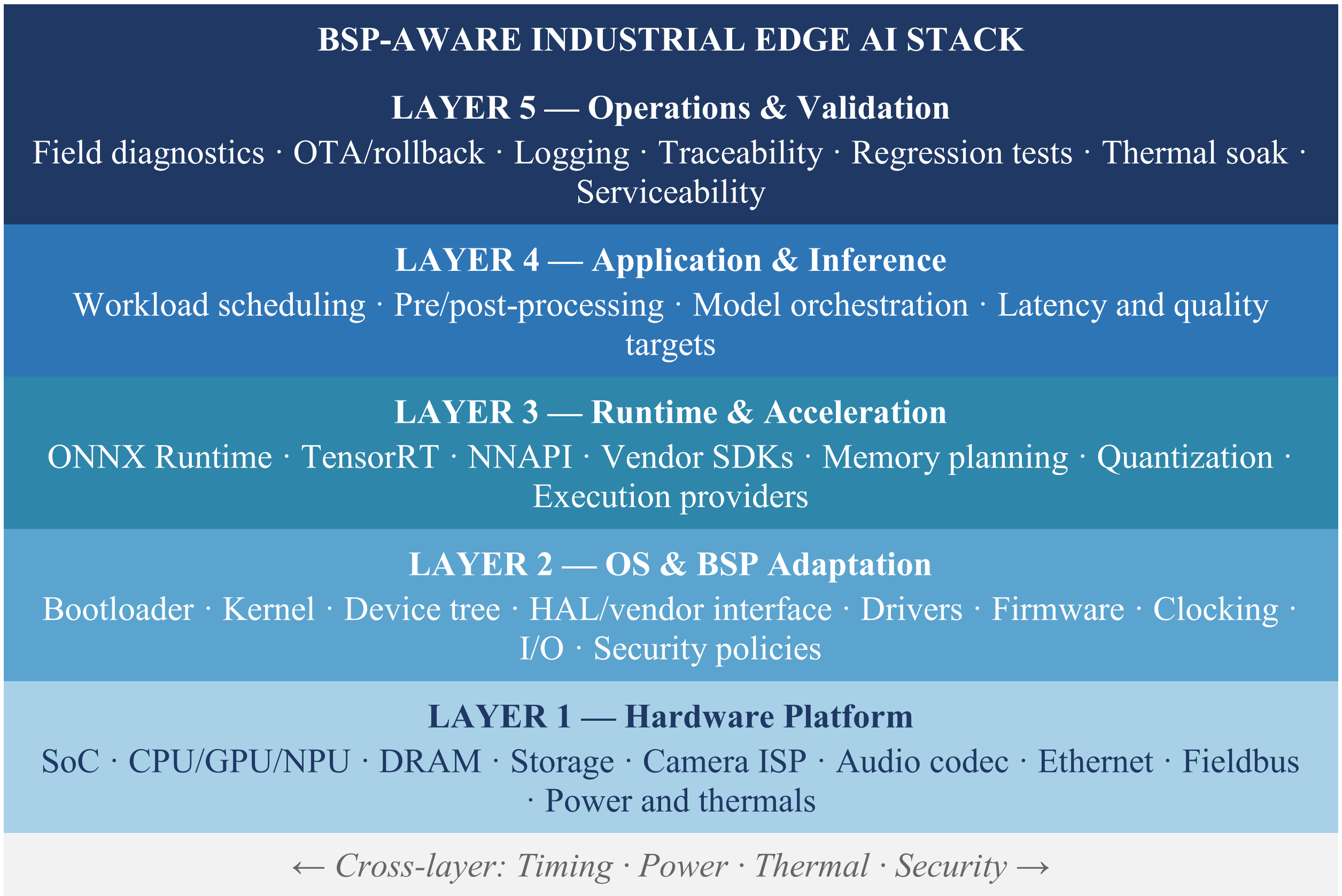

*Figure 1: BSP-aware industrial Edge AI stack. AI behavior is shaped by multiple layers below the application, while timing, power, thermal management, and security cut across the full stack.*

### B. Layer 1: Hardware Platform

The bottom layer contains the SoC, accelerator blocks, memory hierarchy, sensor interfaces, storage, power delivery, and thermal envelope. This is where the first deployment decisions should be made, because later software choices cannot remove physical mismatches. A board selected for demo convenience may later be found to lack the required camera interface bandwidth, deterministic I/O, ECC requirements, service connectors, or thermal headroom for sustained inference.

The correct hardware question is therefore not merely "can this board run the model?" but rather "can this platform run the full workload under the actual duty cycle with the required I/O set, service strategy, and lifecycle support?" That includes development factors such as availability of debug access, documentation quality, and maturity of the vendor software stack. If these are weak, the engineering effort shifts from product development to platform rescue.

### C. Layer 2: BSP and OS Adaptation

This layer translates the board into a usable execution platform. It includes bootloaders, kernel configuration, device tree, peripheral drivers, firmware loading, security settings, HALs or vendor interfaces, and the core system image. In Android, HALs provide the abstraction boundary that allows higher layers to remain agnostic to lower-level driver implementations [1]. In Linux-based systems, the equivalent role is distributed across kernel drivers, user-space services, and vendor libraries.

For Edge AI, this layer must be treated as part of the inference path. A camera HAL is not peripheral to a vision model; it defines how frames enter the system. An audio driver is not separate from a wake-word pipeline; it shapes the signal statistics. Network stack behavior is not separate from inference scheduling when time alignment and packet arrival determine which window of data is analyzed. If this layer is unstable, the application inherits nondeterminism that is extremely expensive to debug later.

### D. Layer 3: Runtime and Acceleration

The third layer contains the software used to execute the model efficiently: ONNX Runtime, TensorRT, NNAPI, vendor NPUs, media frameworks, and any conversion or build tools required to reach the preferred execution path. ONNX Runtime explicitly describes itself as a cross-platform model accelerator with interfaces for integrating hardware-specific libraries [5]. TensorRT describes its role as optimizing and accelerating deep learning inference on NVIDIA GPUs, supporting precision modes, dynamic shapes, and deployment-specific optimizations [6]. These capabilities are valuable, but they only behave as expected when the lower platform layers expose the necessary devices, memory behavior, and driver support.

This is a crucial system point: the runtime does not float above the platform. It is a negotiation between model graph characteristics and platform capabilities. Teams often discover late that an execution provider is present but not fully usable, that a fallback path silently shifts work to the CPU, or that a dynamic-shape model causes unacceptable allocator churn. A BSP-aware process surfaces those mismatches during runtime integration rather than after customer testing begins.

### E. Layer 4: Application and Inference Logic

At this layer live the familiar AI artifacts: preprocessing, postprocessing, orchestration, decision logic, UI integration, and service interfaces. This is also the layer where most program managers first seek visible progress, because it produces demonstrable behavior. The problem is not that the application layer matters too little; it is that it is often asked to compensate for unresolved platform problems below it.

In a systems-first process, application engineering remains important but is intentionally sequenced after platform evidence exists. For example, a vision pipeline should not finalize threshold logic until camera path stability, color processing, and sustained frame cadence are characterized. Likewise, a predictive maintenance pipeline should not freeze anomaly thresholds until the audio or vibration acquisition path is confirmed under the real sampling configuration. The application becomes more robust when it is built on measured platform behavior rather than on assumptions.

### F. Layer 5: Operations and Validation

The top layer includes field diagnostics, logging, observability, OTA strategy, rollback support, traceability, and regression qualification. Industrial deployments live or die here because the question eventually becomes not "did the model run once?" but "can the fleet be maintained safely and diagnosed quickly under change?" This layer also connects software engineering with operations engineering. A system that cannot report accelerator fallback, sensor-drop statistics, thermal throttling events, or update provenance is difficult to trust no matter how accurate the model was in the lab.

**TABLE I: Framework Layers, Engineering Intent, and Minimum Evidence Artifacts**

| Layer | Primary Goal | What Must Be True Before Moving Up | Evidence Artifact Expected |
|---|---|---|---|
| **Hardware** | Fit workload to a real product platform | Required I/O, compute, power, storage, and thermal envelope are feasible for the duty cycle | Platform selection note, interface map, thermal estimate, lifecycle/support statement, carrier-board revision control |
| **BSP / OS** | Make the board reproducibly bootable and hardware-accessible | Boot chain, kernel, device tree, drivers, and vendor interfaces expose all required peripherals consistently | Boot logs, device-tree diff history, driver enablement record, peripheral smoke-test report, recovery path validation |
| **Runtime / Acceleration** | Ensure the model runs on the intended execution path | Model conversion, execution provider selection, precision mode, and memory behavior are verified on target | Runtime capability matrix, accelerator verification log, latency trace, fallback detection record, model package manifest |
| **Application / Inference** | Deliver correct task logic under real device conditions | Pre/post-processing, scheduling, thresholds, and data plumbing are tuned to actual platform behavior | End-to-end validation set report, task-level KPI record, error-budget note, service API contract |
| **Operations / Validation** | Make the device supportable over its field life | Update, rollback, logging, monitoring, and soak tests prove the system remains diagnosable and recoverable | Qualification plan, OTA procedure, rollback test result, observability schema, traceability matrix |

## IV. DEPLOYMENT LIFECYCLE AND VALIDATION GATES

### A. Why a Gated Lifecycle Is Necessary

In model-first projects, validation is frequently postponed until the end, at which point failures appear as long bug lists without a unifying structure. A systems approach replaces that pattern with gates. A gate is not bureaucracy; it is a deliberate checkpoint proving that the next layer will not be built on uncertain assumptions. Figure 2 illustrates this sequence.

| **PHASE 1 Platform Selection** SoC · Interfaces Lifecycle · Toolchain | **PHASE 2 BSP Bring-Up** Boot chain · Kernel Device tree | **PHASE 3 Peripheral Enablement** Camera · Audio Network · GPIO | **PHASE 4 Runtime Integration** Accelerators · APKs Model packaging | **PHASE 5 Qualification** Stress · Thermal EMC · Recovery · OTA |
|---|---|---|---|---|
| | **GATE A** Boots reproducibly Exposes debug path | **GATE B** All required I/O validated with traces | **GATE C** Runtime matches accelerator target | **GATE D** Field recovery & sustained load pass |

*Figure 2: Deployment lifecycle with explicit validation gates. The gates prevent teams from carrying platform uncertainty upward into runtime, application, and qualification work.*

### B. Phase 1: Platform Selection and Compatibility Screening

The first phase combines system requirements with platform due diligence. The team should confirm the interfaces required by the actual product, not by the demo. That includes sensor count and type, display needs, storage endurance, industrial buses, debug access, expected duty cycle, and any security or regulatory constraints. At the same time, the team should examine the software ecosystem: BSP cadence, kernel branch policy, vendor documentation quality, reference designs, accelerator support, and availability of long-term maintenance.

This phase is where many future problems can be prevented cheaply. For example, a platform may appear attractive based on TOPS or benchmark headlines but may depend on a young BSP branch, incomplete camera enablement, or unclear upgrade compatibility. AOSP's reference-board guidance notes that the BSP for a reference board is typically obtained from the board manufacturer rather than from AOSP itself [2]. That is a reminder that deployability is vendor- and board-specific, not guaranteed by the application framework alone.

### C. Phase 2: BSP Bring-Up and Boot Integrity

Once the platform is selected, the next job is to prove that the board boots reproducibly and remains recoverable. The output of this phase is not just a successful boot; it is a dependable boot path with observability. Engineering teams should capture serial logs, confirm storage partitioning, validate watchdog behavior, test failure recovery, and document the exact software provenance. NXP's documentation is valuable here because it treats bootloader, Linux image creation, and root filesystem generation as first-class BSP tasks rather than as invisible preconditions [3].

A production deployment should leave this phase only when the platform can be brought back from common failure scenarios such as interrupted flashing, configuration drift, or partially failed updates. Without that guarantee, later AI work may appear to progress while operational risk silently accumulates.

### D. Phase 3: Peripheral Enablement and I/O Characterization

This phase is often underestimated. It includes not only making camera, audio, network, USB, GPIO, and storage devices "work," but characterizing how they work. Is frame cadence stable? Are timestamps monotonic and synchronized? Does the codec path preserve expected amplitude and sample rate? Are Ethernet and time-sync settings adequate for multi-device correlation? What happens under simultaneous traffic and inference load? These questions determine the statistical quality of the model input and the operational integrity of the device.

A team should not proceed to application tuning until these measurements are recorded. Otherwise, application behavior gets shaped by invisible platform variability. The result is a brittle system whose thresholds must be revisited every time the board revision, driver version, or firmware configuration changes.

### E. Phase 4: Runtime Integration and Model Packaging

The runtime phase is where model conversion, quantization, graph optimization, and execution-provider selection meet the target hardware. The literature on embedded AI tooling shows why this phase deserves explicit structure. Hasanpour et al. describe embedded AI deployment as a multi-step workflow involving model generation, optimization, conversion, and deployment, and emphasize that the process becomes burdensome when many models and devices must be managed together [7]. In practice, industrial teams must also verify that the chosen runtime actually exercises the desired accelerator path, that dynamic-shape behavior is acceptable, and that memory allocation remains stable during long-running sessions.

This phase is where silent fallback must be treated as a defect. If a runtime is nominally installed but certain operators or shapes cause execution to fall back to the CPU, the system may still appear functional in low-volume testing while violating latency or thermal expectations in the field. A BSP-aware process therefore requires targeted verification rather than assuming that installation equals acceleration.

### F. Phase 5: Qualification Under Industrial Conditions

Final qualification is not a single regression pass. It is a campaign that combines functional tests with environmental and operational stress. That usually includes soak tests, repeated boot and recovery cycles, storage-stress scenarios, power interruption tests, update and rollback drills, and multi-hour or multi-day workload execution with telemetry enabled. When relevant, it should also include time synchronization validation, network-loss scenarios, and degradation behavior when an accelerator is unavailable.

At this phase, the system must be evaluated as a product, not as a development board running a model. If a device cannot be updated safely, traced to a software bill of materials, or diagnosed remotely, the deployment is still incomplete even if the model output looks correct.

**TABLE II: Validation Gates and Exit Criteria**

| Gate | Where It Sits | Minimum Exit Criteria | Why It Matters for Edge AI |
|---|---|---|---|
| A | After bring-up | Reproducible boot, serial/debug access, storage map verified, recovery path documented, watchdog behavior known | Prevents later AI integration from masking fundamental board instability |
| B | After peripheral enablement | All required I/O paths validated with representative data, time sync and buffering characterized, error paths observable | Ensures model inputs and event timing reflect the real platform rather than lab assumptions |
| C | After runtime integration | Intended accelerator path confirmed, fallback detection in place, sustained memory behavior acceptable, packaging reproducible | Avoids surprise CPU fallback, thermal overload, or non-reproducible model execution |
| D | Before production release | OTA and rollback proven, soak tests passed, power-loss and restart behavior acceptable, observability hooks complete | Converts a working demo into a maintainable industrial device |

## V. CROSS-LAYER FAILURE MODES AND DIAGNOSTIC STRATEGY

### A. From Symptom-Based to Chain-of-Causality Debugging

When an application-first process encounters trouble, debugging often begins where the symptom appears: inaccurate detections, missed wake events, intermittent latency spikes, or random service restarts. The danger is that the symptom is frequently two or three layers removed from the root cause. Figure 3 illustrates this propagation. The engineering response should therefore be chain-of-causality debugging, where each symptom is traced backward through runtime, OS, BSP, and physical I/O.

**Figure 3: Cross-Layer Failure Propagation in Industrial Edge AI Systems**

| Physical / Input Domain | BSP / OS Manifestation | Inference Symptom | Production Impact |
|---|---|---|---|
| Camera · Sensor timing · ISP tuning · Illumination | Frame drops · Color mismatch · Buffer starvation | Unstable logits · False rejects · Validation drift | Scrap increase · Operator distrust · Re-qualification |
| Audio codec gain · Clock / DMA jitter | Underrun · Clipping · Incorrect sample rate | Wake-word miss · Poor anomaly scoring | Missed events · Unsafe silence |
| Network link flaps · PTP / timestamp skew | Packet loss · Stale queues · Wrong time ordering | Late or duplicate inference · Event correlation errors | Bad root-cause analysis · Service calls |

*Figure 3: A system defect often begins in the physical or BSP layer and only appears to the user as an application or AI-quality problem.*

### B. Vision Path Failures

Vision systems are particularly sensitive because the AI model sees the combined output of optics, sensor timing, ISP configuration, memory movement, and preprocessing. A color-space mismatch, a change in exposure policy, a hidden resize, or a dropped-frame pattern can all alter the tensor stream even if the application code remains untouched. The work by Cidon et al. on edge-device instability underscores this point by showing that image signal processing differences can materially affect model outcomes [8]. In industrial inspection systems, the practical implication is that camera acceptance must include signal-path characterization, not merely camera detection.

Typical diagnostics include raw-frame capture, metadata logging, cadence traces, dropped-frame counters, buffer ownership analysis, and controlled comparison between training-time preprocessing and deployment-time preprocessing. The goal is to determine whether the model is failing on the data or whether the data path is failing the model.

### C. Audio and Low-Bandwidth Sensor Failures

Audio and vibration pipelines often look simple on architecture diagrams but fail in subtle ways. Sample-rate misconfiguration, gain drift, clipping, and DMA jitter can degrade task performance while keeping the system nominally "alive." Such failures are especially dangerous in anomaly detection, acoustic classification, or speech-triggered systems because they tend to produce variable false-negative behavior rather than obvious crashes.

The correct diagnostic set includes loopback tests, known-tone injection, window-alignment checks, buffer underrun counters, and confirmation that the deployment preprocessing precisely matches the training configuration. A system that only passes subjective listening tests is not sufficiently characterized for production AI.

### D. Network, Time, and Fleet-Management Failures

In connected industrial systems, network behavior is often treated as an IT problem rather than an AI problem. That separation becomes invalid when multiple streams must be correlated, when inference decisions are timestamped, or when cloud and edge components cooperate. Packet loss, time drift, queue buildup, and delayed persistence can all make a correct model appear inconsistent because its outputs are compared against the wrong time window or delivered out of order.

Diagnostic discipline here includes packet capture, queue-depth metrics, synchronization verification, reboot persistence checks, and explicit validation of behavior under link interruption. The deployment should also specify what happens when connectivity degrades: does the system continue locally, buffer, discard, or change decision confidence? Systems thinking forces these policies to be defined in advance.

### E. Thermal, Power, and Storage Failures

Thermal throttling, brownouts, and storage pressure are common reasons why an apparently successful demo degrades in the field. DeepEdgeBench shows why sustained edge behavior deserves attention by comparing inference time and power behavior across several resource-constrained devices and accelerator configurations [9]. The takeaway is not that one board wins universally, but that throughput and power behavior are highly platform-dependent.

Storage and update behavior are similarly underestimated. Model packages, logs, and temporary artifacts can create write amplification or exhaust space on embedded media. A systems-first deployment therefore treats storage, update cadence, and rollback policy as part of the AI delivery path rather than as generic platform housekeeping.

**TABLE III: Common Failure Modes, Likely Root Causes, and Preferred Diagnostic Artifacts**

| Observed Symptom | Likely Root Cause Domain | Typical Root Causes | Diagnostic Artifacts |
|---|---|---|---|
| **Intermittent false rejects in vision task** | Sensor / ISP / preprocessing path | Exposure drift, color conversion mismatch, frame drops, hidden resize, metadata mismatch | Raw frame dump, ISP settings snapshot, frame-timestamp log, preprocessing parity test, dropped-frame counters |
| **Wake-word misses or unstable anomaly scores** | Audio codec / DMA / sampling path | Clipping, wrong gain staging, sample-rate conversion, buffer underruns, window misalignment | Known-tone injection trace, codec register dump, loopback record, underrun statistics, preprocessing checksum |
| **Latency spikes after a few minutes of runtime** | Runtime / memory / thermal policy | CPU fallback, allocator churn, thermal throttling, concurrent I/O contention | Accelerator verification log, perf trace, thermal telemetry, memory allocation trace, scheduler timeline |
| **Good lab behavior but poor field reproducibility** | Mixed fleet / BSP version drift | Board revision mismatch, driver updates, vendor-library incompatibility, environmental variation | Software bill of materials, board-revision matrix, BSP release manifest, field telemetry comparison, rollback experiment |
| **Model runs, but update is unsafe** | Boot / storage / OTA design | Incomplete A/B strategy, bootloader state issue, insufficient free space, rollback not validated | Partition map, update transaction log, power-loss test result, recovery boot capture, storage endurance note |
| **Incorrect event ordering across devices** | Network / time sync | PTP failure, NTP drift, queue backlog, out-of-order persistence | Packet capture, queue-depth trace, sync-offset log, timestamp source audit |

## VI. COMPARATIVE ANALYSIS: APPLICATION-FIRST VS. SYSTEMS-FIRST DEPLOYMENT

### A. Why This Comparison Is Important

Industrial teams rarely choose between two fully documented methods. More often, they drift into an application-first workflow because that is the easiest path to a visible demo. The result may look efficient at the beginning, because UI, inference, and business logic can be shown quickly. The hidden cost is that unresolved platform issues accumulate until late validation, at which point every bug is harder to localize and every schedule estimate becomes fragile.

A systems-first workflow appears slower because it spends earlier effort on bring-up, device characterization, and validation gates. Yet this work buys something extremely valuable: reduction of uncertainty. The comparison therefore should not be framed as "fast vs. slow," but as "early visible output vs. controlled integration risk." Table IV summarizes this contrast.

**TABLE IV: Application-First and Systems-First Deployment Patterns Compared**

| Dimension | Application-First Pattern | Systems-First, BSP-Aware Pattern |
|---|---|---|
| **Starting assumption** | If the model runs on the target once, the remaining work is packaging and tuning | The model is one element in a larger execution chain whose lower layers must be characterized first |
| **Early project output** | Fast demos, visible UI progress, quick proof-of-concept behavior | Slower visible demo velocity, but stronger evidence about boot, I/O, recovery, and runtime compatibility |
| **Debugging style** | Symptom-led; late defects appear as app bugs, threshold issues, or unexplained performance regressions | Chain-of-causality; failures are mapped backward through runtime, OS, BSP, and hardware |
| **Risk profile** | Hidden integration debt accumulates and tends to surface late, when schedule pressure is highest | More work happens earlier, but uncertainty is reduced before application assumptions harden |
| **Adaptation to mixed fleets** | Often weak; board-specific differences are discovered late and patched ad hoc | Stronger; BSP versioning, board revision control, and validation evidence are planned as first-class artifacts |
| **Field serviceability** | Frequently underdefined until release pressure forces simplified update procedures | Explicitly designed through OTA, rollback, logging, provenance, and recovery validation |
| **Long-run reliability** | Sensitive to thermal drift, hidden fallback paths, and unmeasured I/O variation | More resilient because platform behavior is qualified under representative operating conditions |
| **Program-management consequence** | Appears fast until defect localization becomes difficult and rework cascades across teams | Appears methodical from the start and supports more credible schedule and risk communication |

### B. Observed Results from the Evidence Base

The literature and vendor documentation reviewed for this paper support four observed results.

Observed result 1: device-side variation is not marginal. Cidon et al. showed that lower-layer differences, including image signal processing, can produce nontrivial prediction divergence across edge devices [8]. This means a model's offline metric is not a complete deployment guarantee.

Observed result 2: deployment is inherently multi-stage. EdgeMark formalizes embedded AI deployment as a sequence of generation, optimization, conversion, and deployment steps [7]. Industrial deployments add platform qualification, recovery validation, and fleet operations on top of that sequence.

Observed result 3: heterogeneous and aging fleets are normal in edge systems. FLEdge documents the challenge of hardware diversity and notes that deployed embedded hardware can remain in service for more than five years in reliability-sensitive systems [10]. In practice, this invalidates any workflow that assumes a single fresh board target.

Observed result 4: acceleration only matters when the platform really exposes it. Vendor runtime documents emphasize integration with hardware-specific libraries and accelerators [5], [6]. The implication is obvious but often ignored: acceleration benefits are contingent on correct driver, kernel, and vendor-library alignment.

### C. Detailed Conclusion of the Comparison

The most important comparison outcome is not that the systems-first path eliminates defects. No realistic engineering method can do that. The stronger conclusion is that systems-first work changes when uncertainty is discovered and where the organization is able to act on it. In an application-first process, uncertainty remains hidden inside the platform until the application is already expected to behave consistently. In a systems-first process, the platform is interrogated early enough that assumptions about thresholds, runtime behavior, latency budgets, and service workflows are built on evidence.

That shift has a direct effect on program behavior. Design reviews become more concrete because they can reference boot logs, peripheral traces, runtime verification records, and thermal telemetry instead of speculation. Integration meetings become less adversarial because defects can be localized to a layer with evidence. Product-management communication improves because risks are tied to gates rather than vague readiness claims. For industrial companies, that improvement in engineering clarity is often more valuable than a small early increase in demo velocity.

## VII. INDUSTRIAL USE-CASE MAPPING

**TABLE V: Representative Industrial Use Cases and BSP-Sensitive Deployment Concerns**

| Use Case | Dominant Technical Objective | BSP-Sensitive Concerns | Validation Focus |
|---|---|---|---|
| **Inline visual inspection** | Stable detection or classification under production lighting and cadence | Camera driver maturity, ISP tuning, trigger timing, memory bandwidth, sustained accelerator use | Frame-path characterization, metadata parity, false-reject analysis, thermal soak under full line rate |
| **Mobile robot / AGV perception** | Low-latency, time-consistent perception and decision support | Multi-sensor timestamp integrity, network quality, degraded-mode behavior, power variation | Sync accuracy, concurrency stress, mobility-induced thermal and power testing, fallback-mode validation |
| **Predictive maintenance gateway** | Long-run anomaly detection from sampled data streams | Codec or ADC configuration, DMA stability, storage persistence, background-work interference | Signal-path parity testing, extended soak, reboot persistence, queue and buffering analysis |
| **Industrial terminal with local AI** | Balance UI responsiveness with dependable local inference | Graphics and inference contention, camera path coexistence, multimedia stack maturity, update safety | Mixed-workload profiling, user-perceived latency, storage/update drills, sustained thermal observation |

### A. Machine Vision Inspection Nodes

Vision inspection systems are one of the clearest examples of why BSP awareness matters. They are often evaluated primarily by model accuracy, yet the deployed quality envelope is usually dominated by camera integration details: sensor timing, trigger behavior, exposure control, ISP tuning, memory bandwidth, and sustained thermal behavior. If any of these are unstable, the vision model becomes a convenient but misleading place to blame failures.

A robust machine vision node therefore starts with camera-path characterization, not with threshold polishing. The BSP-aware framework is especially useful here because it makes the camera pipeline, frame transport path, and accelerator path visible as explicit engineering layers. Teams should capture reference raw frames, preserve metadata, validate timing, and freeze the preprocessing contract before finalizing decision logic.

### B. Autonomous Mobile Robots and Mobile Industrial Platforms

Mobile platforms add motion, multiple sensors, intermittent connectivity, and power variation. Their Edge AI stack is therefore shaped not only by accelerator performance but also by synchronization, sensor fusion consistency, and degraded-mode behavior. A systems mindset is essential because perception errors may originate from dropped frames, timestamp skew, or transport contention rather than from the model graph itself.

On such systems, the BSP-aware framework helps engineering teams decide where deterministic behavior is required and where graceful degradation is acceptable. For example, it is often preferable to define a lower-throughput but stable fallback mode rather than leave the system to silently oscillate between accelerator and CPU execution.

### C. Predictive Maintenance Gateways

Predictive maintenance devices frequently appear simpler than robot or vision systems because they process lower-bandwidth signals such as vibration, current, or acoustic data. In practice, they demand discipline in sampling, windowing, and long-duration reliability. The model's quality depends on acquisition integrity and time alignment just as much as on feature engineering.

A BSP-aware process ensures that signal capture, sample format, DMA behavior, and persistence policy are validated before anomaly thresholds are tuned. This is especially important when the gateway also performs communications, local logging, and update handling, because those workloads can interfere with real-time acquisition if not characterized.

### D. HMI-Plus-AI Industrial Terminals

Industrial HMIs that add local AI inference must reconcile display workloads, multimedia pipelines, user interaction, and inference execution on the same platform. The platform can appear powerful in short demos but fail under sustained mixed workload if graphics, camera, and inference contend for memory bandwidth or thermal headroom.

In these products, the advantage of a systems mindset is that UI performance and AI performance are evaluated together rather than in separate lab silos. The deployment decision becomes a full product decision, not an isolated model decision.

**Figure 4: Recommended Validation Emphasis Across Deployment Stages**

| Concern Area | Bring-Up | Driver Validation | Runtime Integration | System Qualification |
|---|---|---|---|---|
| Boot/Recovery | Low | Low | Med | HIGH |
| Camera | Low | Med | Med | Med |
| Audio | Low | Med | Med | Med |
| Network / PTP | Low | Med | Med | HIGH |
| Storage / OTA | Med | Low | Med | HIGH |
| Thermal Soak | Low | Low | Med | HIGH |
| Power Interrupt | Med | Low | Low | HIGH |

*Figure 4: High-emphasis cells indicate where engineering effort most strongly reduces later deployment risk.*

## IMPLEMENTATION CHECKLIST FOR ENGINEERING TEAMS

### A. What a Real Deployment Packet Should Contain

A BSP-aware deployment is easier to govern when each program maintains a minimal evidence packet. That packet should include at least the following: exact BSP revision, board revision, bootloader and kernel provenance, device-tree differences, peripheral enablement report, runtime capability matrix, accelerator verification logs, model package manifest, validation-set record, OTA/rollback procedure, and soak-test telemetry summary. These are not paperwork for their own sake. They are the fastest way to keep a field problem from turning into a weeks-long archaeology project.

### B. Checklist Before Production Sign-Off

Before production sign-off, an engineering review should be able to answer the following questions with evidence rather than optimism:

5. Is the exact board and BSP combination frozen and traceable?
6. Can the system boot, recover, and roll back safely after failed updates or power interruption?
7. Have all required sensor and I/O paths been validated with representative data and timing?
8. Has the intended acceleration path been positively verified, including detection of CPU fallback?
9. Do the deployed preprocessing and postprocessing steps match the validated model contract?
10. Has sustained behavior been characterized under realistic thermal, storage, and concurrency conditions?
11. Are field diagnostics sufficient to determine whether a defect belongs to hardware, BSP, runtime, or application logic?

> **NOTE**
>
> If any of these answers is "not yet," the correct interpretation is not that the system is almost ready. It is that the system remains under-specified in a production-relevant dimension.

## VIII. DISCUSSION

### A. Relationship to MLOps and Platform Engineering

Some organizations attempt to solve embedded deployment risk by importing cloud-style MLOps practices alone. While CI/CD, model registries, and experiment tracking are useful, they do not replace platform engineering. A CI pipeline cannot compensate for an unstable sensor clock, an immature camera HAL, or unverified fallback behavior on the target device. The right view is that embedded MLOps must sit on top of a qualified platform substrate. In industrial environments, platform engineering and MLOps are complements, not substitutes.

### B. What This Framework Does Not Claim

This paper does not claim that one runtime, one SoC, or one OS is universally best. It also does not claim that every project needs the same depth of validation. The correct validation depth depends on product criticality, duty cycle, field service model, and fleet size. What the paper does claim, based on the reviewed evidence, is that the dominant sources of deployment failure are often cross-layer and therefore cannot be managed by model-centric thinking alone.

### C. How Teams Should Adapt the Framework

The framework should be scaled to the product. For a single-purpose sensor node, the evidence packet may be smaller, but the gates should still exist. For a complex Android-based industrial terminal, the gates may expand to include HAL compatibility, boot-time optimization, and graphics-inference coexistence. What should not change is the principle that every layer must produce objective evidence before the next layer assumes stability.

## IX. CONCLUSION

### A. Detailed Synthesis of the Observed Results

The reviewed evidence leads to a clear conclusion: Edge AI deployment quality is strongly shaped by system behavior below the model. Android, NXP, and NVIDIA all document platform stacks in which hardware access and customization are mediated by BSP-level components such as HALs, kernels, bootloaders, root filesystems, and vendor interfaces [1], [3], [4]. That architectural fact alone should discourage any engineering plan that treats the BSP as an afterthought.

The systems literature sharpens the point. Cidon et al. show that model outputs can diverge materially across devices because of lower-layer behavior such as image signal processing [8]. Hasanpour et al. show that embedded AI deployment is not a single action but a workflow containing generation, optimization, conversion, and deployment steps that benefit from automation and reproducibility [7]. Woisetschläger et al. show that heterogeneous, aging, reliability-constrained edge fleets are a normal deployment reality rather than an exception [10]. DeepEdgeBench shows that inference and power behavior vary substantially by device and framework configuration [9]. Considered together, these results support a practical conclusion: the industrial problem is not simply getting a model to run, but getting a complete platform to behave predictably over time.

### B. Comparison Outcome Stated Plainly

Compared with an application-first workflow, a systems-first and BSP-aware workflow produces a more credible path to industrialization. It does not guarantee perfection, and it may appear slower during the first visible weeks of a program. However, it improves the quality of design decisions, makes defect localization easier, reduces the chance of late rework, and creates the evidence needed for safe product release and field maintenance. In real organizations, those advantages typically outweigh the apparent speed of early demos.

The practical difference can be expressed in one sentence. An application-first process asks the platform to adapt to a model that is already assumed to be ready. A systems-first process asks the model and the application to adapt to a platform whose actual behavior has been measured. The second is a better description of how industrial products are successfully built.

### C. Final Recommendation

Teams building industrial Edge AI platforms should adopt a BSP-aware deployment framework from project inception. Platform selection should consider BSP maturity and lifecycle support alongside accelerator capability. Bring-up should be gated by boot and recovery evidence. Peripheral paths should be characterized before model thresholds are frozen. Runtime acceleration should be positively verified rather than assumed. Qualification should include field-operability artifacts such as OTA, rollback, telemetry, and traceability. When these practices are followed, Edge AI stops being a fragile demonstration and becomes a maintainable product capability.